# A Dynamical Systems Perspective on Chimeric Antigen Receptor T-Cell Dosing

Amir A. Toor, MD, [1] Alden Chesney, MD, [2] Jamal Zweit, PhD, [3] Jason Reed, PhD, [4] Shahrukh Hashmi, MD, MPH. [5]

[1] Bone Marrow Transplant Program, Massey Cancer Center, [2] Department of Pathology, [3] Department of Nuclear Medicine, and [4] Department of Physics, Virginia Commonwealth University, Richmond, VA; [5] Division of Hematology, Dept. of Medicine, Mayo Clinic, Rochester, MN & Dept. Of Stem Cell Transplant, Oncology Center, KFSHRC, Saudi Arabia.

Correspondence; Amir A. Toor, MD, Professor of Medicine, Virginia Commonwealth University. E-mail: amir.toor@vcuhealth.org.

~ ~ ~ ~ ~ ~ ~ ~ ~ ~ ~ ~ ~

Chimeric antigen receptor T cells (CAR T cells) are dosed similarly to donor lymphocyte infusions following hematopoietic cell transplantation. In this perspective paper a mathematical basis for personalized dosing of CAR T cells is introduced.

~ ~ ~ ~ ~ ~ ~ ~ ~ ~ ~ ~

Chimeric antigen receptor T cells (CAR-T cells) are genetically engineered T cells transduced with an antibody-like cell surface antigen-binding domain (single chain variable factor, scFv) combined with T cell activating domains. [1] [2] [3] CAR-T cells are highly active against lymphoid malignancies such as acute lymphoblastic leukemia (ALL), non Hodgkins lymphoma (NHL) and multiple myeloma, which express cell surface antigens such as CD19, CD22 and BCMA, mounting a proliferative response with cytotoxicity and clearance of the antigen bearing tumor. [4] This therapy is complicated by severe toxicity, specifically cytokine release syndrome and neurotoxicity. The factors impacting toxicity include CAR T cell dose, tumor burden and potentially T cell subtypes infused. [5] [6] CAR T cell dosing in most trials to date is similar to donor lymphocyte infusion (DLI) dosing following allogeneic hematopoietic cell transplantation (HCT). This dosing schema fails to take into account the unique mechanism of action of CAR T cells, i.e. the relationship between the antigen and transduced, antigen-binding domain driving T cell proliferation and function.

To understand the principle behind DLI dosing, the dynamics of alloreactive T cell responses observed following HCT needs to be understood. The alloreactive donor T cell response in graft-versus-host-diseases (GVHD), is a considerably more complex problem than that of single antigen directed CAR T cell response kinetics. Lymphocyte and T cell proliferation kinetics post HCT demonstrate the following principle features; (a), lymphocytes and T cells follow logistic growth kinetics, with exponential growth and eventual plateau, [7] and (b), T cell clonal repertoire is complex and has a Power law distribution which is maintained after HCT. [8] Based on these findings, whole exome sequencing of HCT donors and recipients was done and identified a multitude of recipient specific polymorphisms, [9] which may yield a large library of potentially alloreactive HLA binding peptides. [10] The distribution of peptide antigen binding affinity to HLA molecules compliments the T cell repertoire in each individual, and as such alloreactive donor T cell repertoire may be simulated if the binding affinities and antigen expression levels are known. [11] [12] These simulations rely on unique T cell clones growing in proportion to the binding affinity of the target alloreactive peptide to the HLA molecule, the affinity of the T cell receptor for the HLA-bound peptide antigen, and the expression levels of the protein from which the peptides are derived. The resulting simulations reproduce patterns seen in clinically observed T cell repertoire. [11] [12]

The principles of antigen-driven T cell proliferative response may be applied to CAR T cells as well; these cells respond to a specific antigen and grow exponentially reaching a peak and then eventually settle to a steady state population in most responding patients. [13] Similar to the case for alloreactive T cell proliferation presented above, the magnitude of the peak response and plateau is likely proportional to the CAR specific antigen expression level (CD19, CD22, BCMA etc.) and the binding affinity of the scFv domain of the CAR construct for the epitope in question, as well as the second signal domain/s included in the CAR construct. [14] [15] Heterogeneity introduced by different binding affinities of the CAR constructs, as well as the target antigen-cell surface expression may result in variable efficacy/toxicity profiles observed with various CAR T cell products, assuming that the toxicity is related to uncontrolled proliferation of CAR T cells and ensuing cytokine storm. Using the above principles T cell proliferation may be modelled as a function of time.

$$N_{t\ (CART)} = \frac{((P.e^{(q-r)t}+1).K_{(CART)}{}^B)*N_{0\ (CART)}}{\left(((P.e^{(q-r)t}+1).K_{(CART)}{}^B)-N_{t-1(CART)}\right)(e^{-rtB})+1} \quad \ldots [1]$$

In this iterating equation the number of CAR-T cells at the beginning of the interaction (time 0) is given as $N_{0\ (CART)}$ (1 in this equation), and the number at time, $t$ is given by $N_{t\ (CART)}$, and for the preceding iteration/time by $N_{t-1\ (CART)}$, and the final steady state level is given by $K_{(CART)}$. The growth rate of CAR-T cells is given by $r$ and the binding affinity of the scFv construct and the antigenic epitope is denoted by $B$. The term $P$ is a multiplier for $K$ and denotes the level of antigen/malignant cell burden in question, with growth rate $q$. The term $e^{(q-r)t}$ computes the change in malignant cell population relative to CAR T cell growth, where if the CAR T cell growth rate $r$ is higher than the growth rate $q$ of the malignant cell population, this term declines and vice versa; the term 1 in this expression accounts for the CAR T cell persistence after the malignant cell burden has been eliminated. CAR T cell growth rate when calculated using equation 1 yields the familiar CAR T cell growth response curve (Figures 1 & 2). [16] [17] [18] [19]

With appropriate parametrization, such a model may be used to establish safe dosing nomograms which allow for the anti-tumor effect of the CAR T cells, to be established while minimizing toxicity making this approach superior to trial and error, phase I type studies. This would require calculating the precise antigen driven CAR T cell proliferation. Consider a simple acute lymphoblastic leukemia (ALL) model; assuming an equilibrium between the malignant cells infiltrating the bone marrow space and in circulation, we may assume that the post infusion CAR T cell population present in blood would be proportional to that present in the marrow, and the circulating CAR T cell numbers would quantitatively reflect the total pool. If the CAR-Ag binding affinity is known, plotting the CAR T cell numbers in blood following an infusion as a function of time by taking repeated measures would allow correlation with disease burden. The CAR T cell growth over time curve and the resulting rate of growth may be correlated with tumor/antigen burden. In such calculations, one may derive a continuous variable representing antigen burden (proportional to tumor load and antigen expression), by determining the absolute blast count in circulation, and in the bone marrow space taking into account cellularity and tumor infiltration expressed as fractions and assuming uniform involvement. Correlation of the rate of CAR T cell growth with antigen burden, and with known CAR-Ag affinity will allow the establishment of dose-response curves for various products. Similar consideration may be applied to studying dose-toxicity relationships with cytokine release syndrome (CRS) and neurotoxicity. Furthermore, unlike DLI where following infusion a mandatory period of several weeks has to elapse before a second infusion may be administered because of concerns for GVHD, such concern does not exist from CAR T cells. This would allow clinical trial to be designed in which multiple low-dose CAR T cell infusions may be tested.

Current CAR T cell dose determining studies are being designed as traditional phase I trials, using cell doses based on donor lymphocyte infusion (DLI) dosing schema.[20] While these have allowed gradual improvement in clinical outcomes, they do not lend themselves to personalized dose determination. These designs are inherently flawed, because CAR T cell and DLI are not comparable therapeutic modalities. With DLI, a multitude of donor T cell clones is infused into the recipient with the expectation that the cognate alloreactive/tumor specific antigen will be present and increasing DLI cell dose will stochastically improve the odds of encountering tumor specific antigens whilst avoiding simultaneously presented normal host antigens. Equation 1 for multiclonal T cells takes the form,[12]

$$N_{t\,(Ti)} = \frac{((P_j.e^{(q-r)t}+1).K_{(Ti)}^{B_j \times Z_i}) * N_{0\,(Ti)}}{\left(((P_j.e^{(q-r)t}+1).K_{(Ti)}^{B_j \times Z_i}) - N_{t-1(Ti)}\right)(e^{-rtB})+1} \dots [2]$$

Equation 2, while similar to equation 1, now depicts the growth of a single, $i^{th}$ T cell clone, when its T cell receptor (TCR) with affinity $Z_i$, ligates the $j^{th}$ peptide bound to an HLA molecule (with an affinity $B_j$). Critically, $P_j$ in this instance depicts the tissue expression of the protein from which the $j^{th}$ peptide is derived, and the cell mass which expresses this protein. This is a significantly smaller value, compared to an epitope on a surface expressed protein which is directly recognized by a CAR, and results in a significantly smaller $K$ for individual T cell clones when compared to CAR T cells. This is because CAR T antigens are expressed at several fold higher levels and available for interactions with CAR T cells, than are HLA-bound-peptides for TCR. Furthermore, the likelihood that tumor associated peptides, and alloreactive peptides will be presented is significantly lower because of the myriad of peptides presented by HLA (Figure 3A). This will render the likelihood of a GVHD-free therapeutic response to DLI, a probability function of the presence of relevant T cell clones (ρT) in the DLI and presentation of the target antigen (ρP) to these clones. The probability of an antitumor response to the DLI (ρR) will in this event be,

$$\rho R = \rho P . \rho T$$

As an example, one may consider that ρT is proportional to the T cell dose infused with the DLI, and is, for example 20% in the lowest dose level of the DLI. If one were to then estimate the ρP to be 20% as it competes with normal tissue peptides also being presented on HLA, the likelihood of a therapeutic response to DLI (without GVHD) will be approximately 4% and will go up with subsequent infusions.

Most other anti-tumor responses will be associated with GVHD. This value is significantly lower than the essentially 100% likelihood of a CAR T cell response (Figure 3B).

In short, unlike DLI, CAR T cells have definite targets generally with high expression, and unlike drugs, these cells grow in response to their target, and this proliferation is proportional to the antigen burden and CAR construct binding affinity. Therefore, CAR T cell dosing needs to account for tumor burden and will in most instances, be a relatively low number of CAR T cells infused to achieve the desired target. Mathematical modelling of the relationship between tumor burden and CAR T cell expansion for various products, may allow for the optimal individualized dosing of CAR T cells for patients.

~ ~ ~ ~ ~ ~ ~ ~ ~ ~ ~ ~

Figure 1. Hypothetical CAR T cell growth kinetics. A. Equation 1 solved with the growth rates, $r$ and $q$ set at 1.5 and 1.2 respectively and $B$ chosen arbitrarily (1/IC50, CD19-CAR ScFv IC50 =1.6 nM). $N_{0\,(CART)}$ is 1, and the constants $P$ and $K$ are 1000 & 1000000 respectively. Initial exponential expansion, followed by steady state CAR T cell population is seen, latter corresponding to a memory T cell population once cytoreduction has occurred.

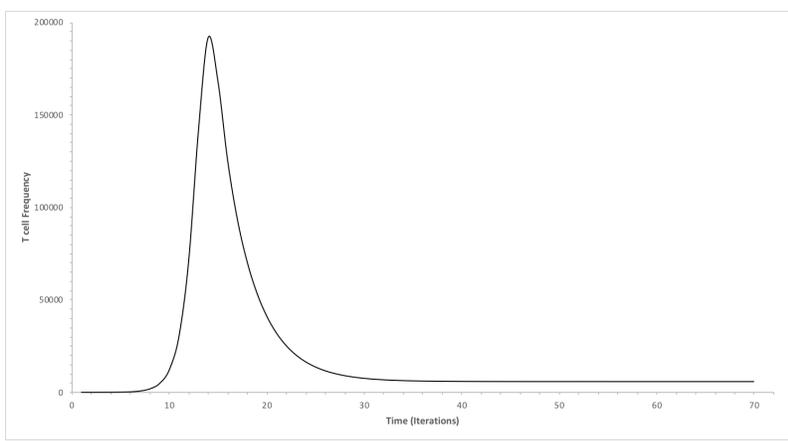

Figure 2. CAR-T cell expansion (black line) and corresponding ALL blast count (red line) decline on logarithmic scale (natural log base 2.7182).

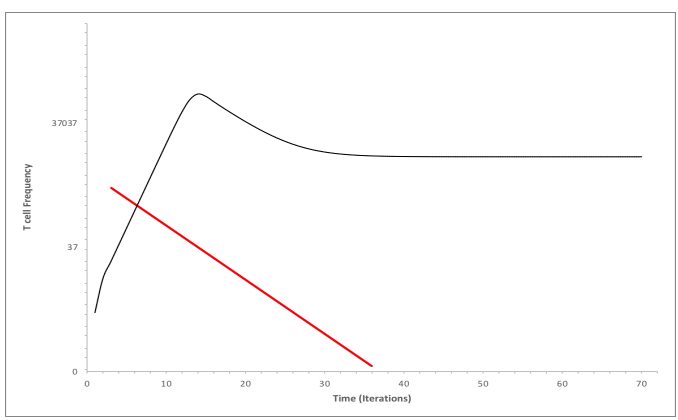

Figure 3. T cell proliferation kinetics. A. HLA-dependent presentation of tumor associated peptides (red) and expansion of donor T cells (green TCR) in response to antigens. There is competing expansion of other non-tumor associated antigen (grey) responsive T cell clones (blue TCR). B. HLA independent expansion of CAR-T cells in response to epitopes on tumor targets.

A.

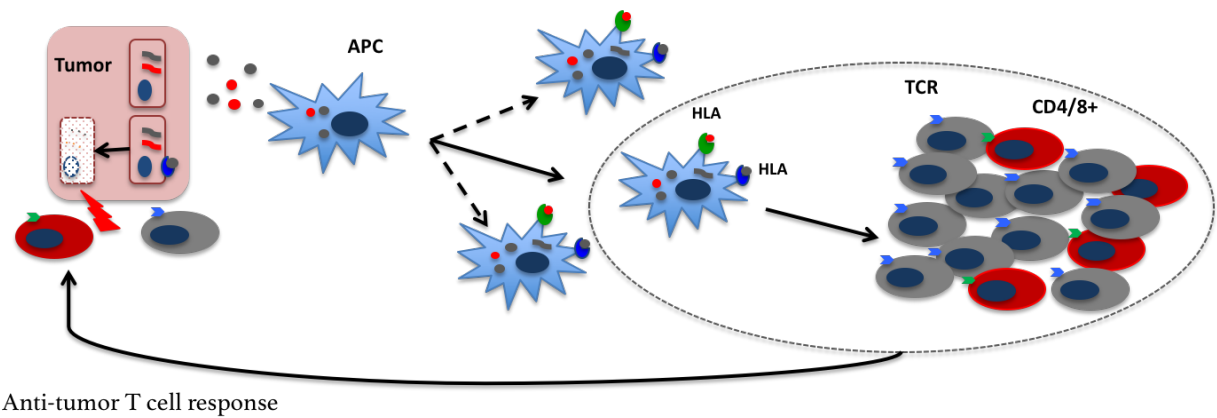

Anti-tumor T cell response

B.

Epitopes on tumor associated cell surface proteins (red) are presented are identified by CAR (grey)

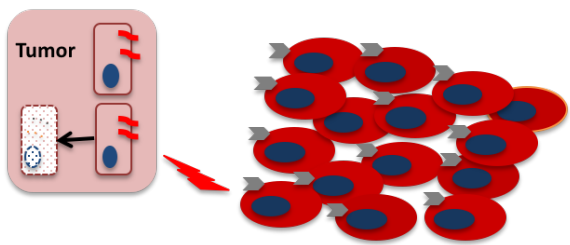

CAR T cell proliferate triggering a CD4+ & CD8+ T cell logistic expansion

Acknowledgements. AAT was supported, in part, by research funding from the NIH-NCI Cancer Center Support Grant (P30-CA016059; PI: Gordon Ginder, MD). Author Contributions. AAT, Developed the idea and wrote the paper. AC, Developed the idea and wrote the paper. JZ, Developed the idea and wrote the paper. JR, Developed the idea and wrote the paper. SKH. Developed the idea and wrote the paper. There are no relevant conflicts of interest to disclose.

References.